\documentclass[preprint,superscriptaddress,showpacs,preprintnumbers,amsmath,amssymb]{revtex4-1}

\usepackage{amsmath}
\usepackage{amssymb}
\usepackage{graphicx}
\usepackage{morefloats}
\usepackage[usenames,dvipsnames]{xcolor}
\usepackage{blkarray}
\usepackage{verbatim}
\usepackage{hyperref}
\usepackage{color}

\begin{document}

\title{Evolutionary Dynamics of Group Formation}

\author{Marco Alberto Javarone}
\email{marcojavarone@gmail.com}
\affiliation{Department of Mathematics and Computer Science, University of Cagliari, Cagliari - Italy}

\author{Daniele Marinazzo}
\email{daniele.marinazzo@ugent.be}
\affiliation{Department of Data Analysis, Faculty of Psychological and Pedagogical Sciences, University of Gent, Gent - Belgium}

\date{\today}

\begin{abstract}
We introduce a model, based on the Evolutionary Game Theory, for studying the dynamics of group formation.
The latter constitutes a relevant phenomenon observed in different animal species, whose individuals tend to cluster together forming groups of different size.
Results of previous investigations suggest that this phenomenon might have similar reasons across different species, such as improving the individual safety (e.g. from predators), and increasing the probability to get food resources. 
Remarkably, the group size might strongly vary from species to species, and sometimes even within the same species.
In the proposed model, an agent population tries to form homogeneous groups. The homogeneity of a group is computed according to a spin vector, that characterizes each agent, and represents a set a features (e.g. physical traits). We analyze the formation of groups of different size, on varying a parameter named 'individual payoff'. The latter represents the gain one agent would receive acting individually. In particular, the agents choose whether to form a group (receiving a 'group payoff'), or if to play individually (receiving an 'individual payoff').
The phase diagram representing the equilibria of our population shows a sharp transition between the 'group phase' and the 'individual phase', in correspondence of a critical 'individual payoff'. In addition, we found that forming (homogeneous) small groups is easier than forming big groups.
To conclude, we deem that our model and the related results supports the hypothesis that the phenomenon of group formation has evolutionary roots.
\end{abstract}
\maketitle
\section{Introduction}
The dynamics of group formation constitutes a topic of interest for a wide number of researchers, spanning from anthropologists to zoologists~\cite{dunbar05,couzin01,couzin07,sumpter01,dunbar04,parrish01,alcock01}, and from social psychologists to economists~\cite{dunbar02,dunbar06,lee01,guillen01,galam01,galam02}.
In general, the formation of a group can be viewed as an emergent phenomenon~\cite{lee01,michod01} where a number of individuals cluster together for performing one or more actions. Accordingly, the lifespan (as well as other characteristics) of a group can vary from case to case, and individuals can change group over time~\cite{stewart01,dunbar07,couzin08}.
In an ecological system, many times being part of a group allows to receive benefits~\cite{dunbar07}, both being a predator and being a prey. For instance, the former can be advantaged during a hunt, e.g. surrounding a prey, while the latter can improve her/his safety staying inside a group~\cite{couzin09}.
Here, we remark that the previous example, referring to predators and preys, can be considered outdated in the case of the human species. However, we should remind that millions of years ago, and maybe even in more recent times, humans have played both roles in their ecosystem.
Different studies suggest that the formation of social groups has evolutionary roots~\cite{dunbar05,couzin05,hofmann01,frewen01,frey01,couzin02,gerard01}, shared among animals belonging to different species. For instance, we can observe groups of fishes (generally named as shoals), of mammalians (named herds or families/tribes in the case of humans), and of birds (named flocks)~\cite{hofmann01}.

\begin{figure}[!ht]
\centering
\includegraphics[width=8cm]{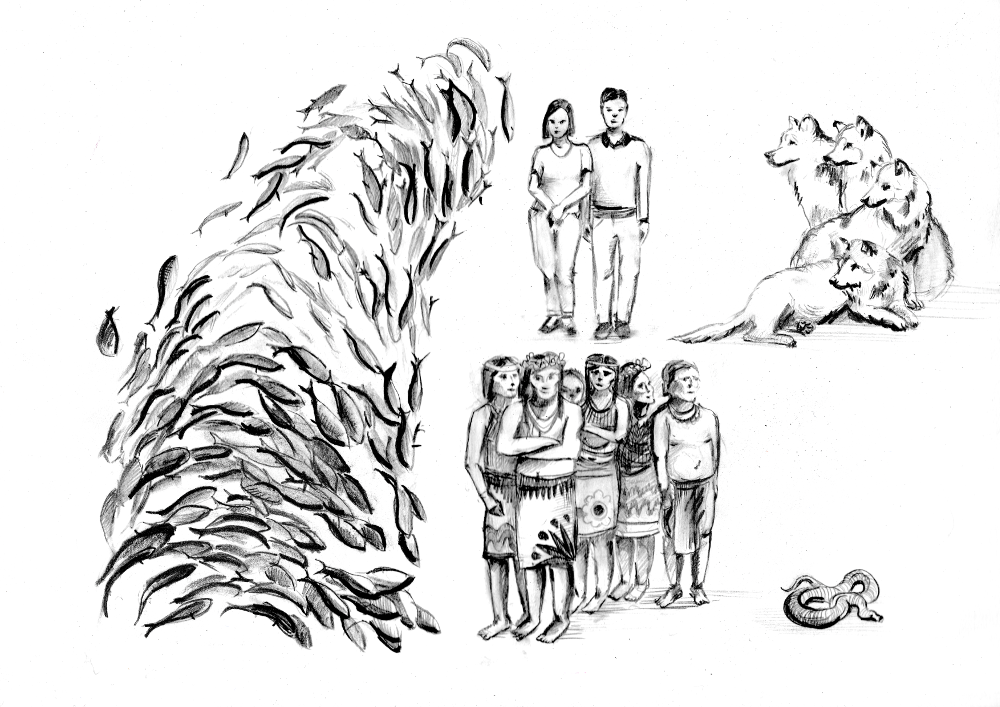}
\caption{Pictorial representation of different animal groups in nature, from the single snake to the shoal of fishes, indicating that the group size does not seem related only to the evolution but also to other factors (to be uncovered). One of the aim of this image is to emphasize that within the same species, as for humans, the average group size may vary according to the living environment and to other conditions (e.g., socio-cultural habits, religious beliefs, and so on). \label{fig:figure_0}}
\end{figure}

What differs, from species to species, is the average size of a group~\cite{dunbar01,couzin03,couzin04,dunbar03,couzin06}, e.g. shoals are usually much bigger than herds, herds are bigger than families, and so on and so forth. In addition, even within the same species, groups of different size can be observed ---see Figure~\ref{fig:figure_0}.
The formation of groups is a phenomenon of interest also beyond the domain of evolutionary biology, as we can mention the formation of sport teams, of business organizations~\cite{guillen01}, and of scientific communities. Even if the motivations that lead to the formation of this kind of groups can be quite different from those that trigger the emergence of groups in nature, in both cases individuals cluster together driven by a rational mindset, i.e. aimed to increase their wealth. 
Therefore, we think that the framework of Evolutionary Game Theory (EGT hereinafter)\cite{perc01,szolnoki01,szolnoki02,perc03,nowak01,nowak02,nowak03,moreno01,santos01,santos02,javarone01,javarone02,antonioni01,grujic01} can be a suitable choice for studying this phenomenon, since it embodies both the rationality and the evolutionary aspect of group formation~\cite{lazaro01,moreno02}. 
When studying the dynamics of group formation, it is important to evaluate the role of similarity. In particular, the heterogeneity of a group can be an advantage, or a disadvantage, depending on the context of reference. Indeed, heterogeneity might refers to different aspects, as physical traits, genetic makeup, or skills. Previous studies (e.g.~\cite{torres01}) reported that social networks show a positive value of assortativity~\cite{newman01}, i.e. it seems individuals be more likely to generate links with their own similar, while other kinds of complex networks~\cite{barabasi01} are more likely to be disassortative (according to an entropic principle~\cite{torres01}).
Thus, in the proposed model, we consider an agent population that forms and breaks groups over time, according to the gain agents receive acting in group or individually. The agent's gain comes from the difference between benefits and costs, in taking a particular action (i.e. group or individual). The gain achieved in group is defined as 'group payoff', while that achieved singularly is defined 'individual payoff'. According to results reported in~\cite{torres01}, here the 'group payoff' is maximized for homogeneous groups.
Results of numerical simulations indicate that for each group size $G$, there is a critical 'individual payoff' between a 'group phase' and an 'individual phase' of the population, i.e. the formation of groups or the individual action.
In addition, forming groups of big size is more difficult than forming small groups.
To conclude, in our view, the achieved results support the hypothesis of an evolutionary mechanism underlying the formation of groups in nature. Notably, we speculate that each animal species has its 'individual payoff', i.e. a kind of gain its individuals receive when they act as single members, and that this parameter might depend also on the considered environment. 
In addition, in the case of human beings, we suppose that the 'individual payoff' might be related also to socio-cultural conditions, leading to the formation of very small groups in the modern civilization, and to the formation of bigger groups (i.e. tribes) in more archaic systems (see~\cite{tehrani01,ray01,dizard01}). Notably, two important differences between the modern civilization and the archaic ones are the living environment and the cultural structure (e.g. relations, laws, etc) of a society, both making a city more suitable than a forest for individual life styles.
The remainder of the paper is organized as follows: Section~\ref{sec:model} introduces the proposed model. Section~\ref{sec:results} shows the results of numerical simulations. Eventually, Section~\ref{sec:conclusions} provides an interpretation of the achieved results, and ends the paper.
\section{The Model}\label{sec:model}
In the proposed model, we consider a population with $N$ agents that can cluster together forming groups of size $G$.
Each agent is represented by a spin vector $S$, of length $L$, e.g. for $L = 6$ the $i$-th agent can be represented as $S_i = [+1,-1,-1,-1,+1,+1]$. Here, each entry of the spin vector can be viewed as a feature, so the homogeneity of a group is measured considering the distance between spin vectors of its members. It is worth to note that we refer to the concept of feature with its more general meaning, since it may vary from species to species. For instance, for many animals (including humans) a feature can be a physical trait, and in the case of humans it can represent also a hobby, or a specific skill, and so on (i.e. not only physical features).
The dynamics of the proposed model is very simple. At each time step a number $G$ of agents, not belonging to any group, is randomly selected. So, selected agents compute the potential payoff they could gain acting together (depending on the homogeneity of the potential group). In particular, the 'group payoff' $\pi_g$ decreases when members have different spin vectors. Then, the value of $\pi_g$ is compared to that of $\pi_i$, i.e. the payoff that agents would gain acting individually.
In doing so, $\pi_i$ and $\pi_g$ are used to compute the probability of forming a group of size $G$, with the selected agents, which reads
\begin{equation}\label{eq:fermi_function}
W(G) = \left(1 + \exp\left[\frac{\pi_g - \pi_i}{K}\right]\right)^{-1}
\end{equation}
\noindent where the constant $K$ parametrizes the uncertainty in taking a decision (i.e. to form, or not, the group). By using $K = 0.5$, we implement a rational approach~\cite{perc03,javarone03}.
After processing a new potential group, the model evaluates if a previous one, randomly selected among those formed at previous time steps, might be broken. The breaking process is performed according to the same equation adopted to generate a group (i.e. Eq.~\ref{eq:fermi_function}).
As mentioned before, the homogeneity of a group is computed according to the spin vector of its members. Accordingly, the group payoff $\pi_g$ is defined as length of the normalized average summation of each spin vector (composing the considered group). In particular, since each entry can be positive (i.e. $+1$) or negative (i.e. $-1$), after computing the average value of a single spin we take its absolute value. So, given spin vectors of length $L$, the 'group payoff' for a group of size $G$ reads
\begin{equation}\label{eq:group_payoff}
\pi_g = \frac{1}{L} \frac{1}{G} \sum_{j=1}^{L} |\sum_{i=1}^{G} v_{ij} |
\end{equation}
\noindent with $v_i$ elements of the spin vector of each agent. Eventually, it is worth noting that the range of $\pi_g$ is $[0,+1]$, while that of the 'individual payoff' $\pi_i$ spans the interval $[-1,+1]$. In doing so, we represent scenarios where acting individually can be very risky (i.e. $\pi_i = -1$), and very convenient (i.e. $\pi_i = +1$). At the same time, we assume that acting in group cannot never lead to a negative payoff. 
Finally, we remark that during each simulation, the value of $\pi_i$ remains constant.
Summarizing, the proposed model can be described as follows:
\begin{enumerate}
\item At $t=0$ generate a population providing each agent with a random spin array;
\item While the number of time step is smaller than $T$:
\item \_\_\_\_ Randomly select $G$ agents, not belonging to other groups;
\item \_\_\_\_ Compute the probability the selected agents form a new group; 
\item \_\_\_\_ Randomly select a group among those previously formed, and compute the probability to break it;
\end{enumerate}
Since we consider an asynchronous dynamics, i.e. only a subset of agents plays at a given time step, the value of $T$ must be big enough in relation to the population size.
\section{Results}\label{sec:results}
\begin{figure*}[!ht]
\centering
\includegraphics[width=16cm]{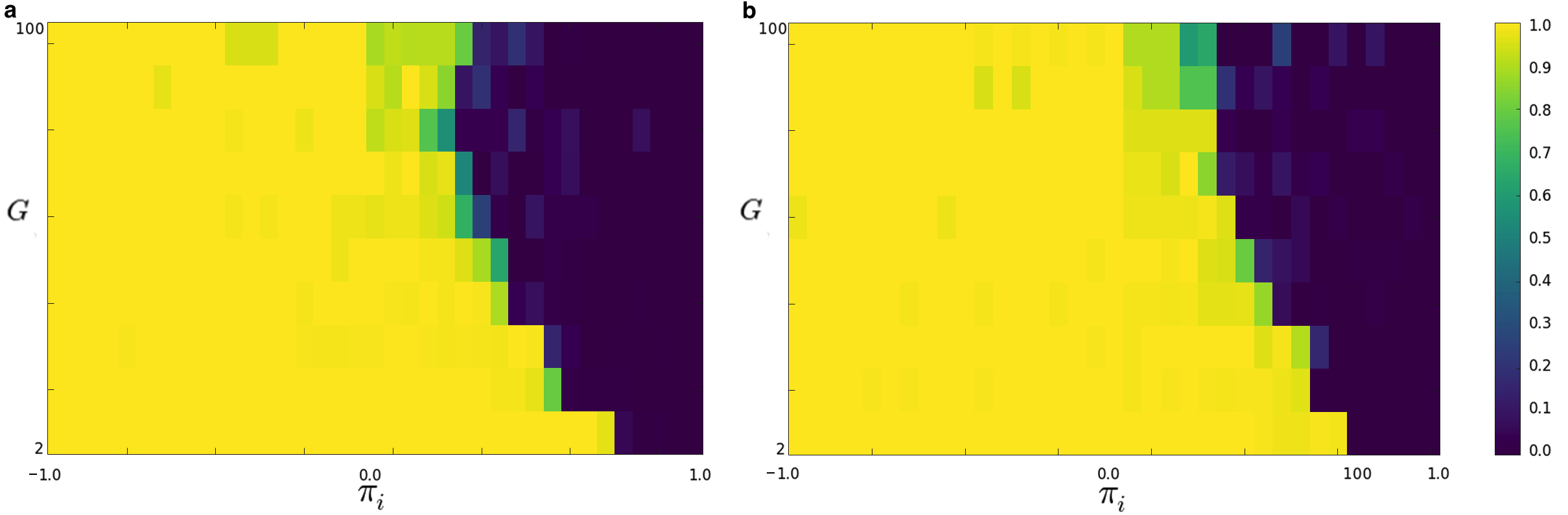}
\caption{Phase diagram of the population, group size $G$ versus the 'individual payoff' $\pi_i$, on varying the length of the spin vector $L$. Yellow indicates the 'group phase', while Blue the 'individual phase'. \textbf{a} $L = 3$ and \textbf{b} $L = 10$. Results have been averaged over different simulation runs. \label{fig:figure_1}}
\end{figure*}

\begin{figure*}[!ht]
\centering
\includegraphics[width=16cm]{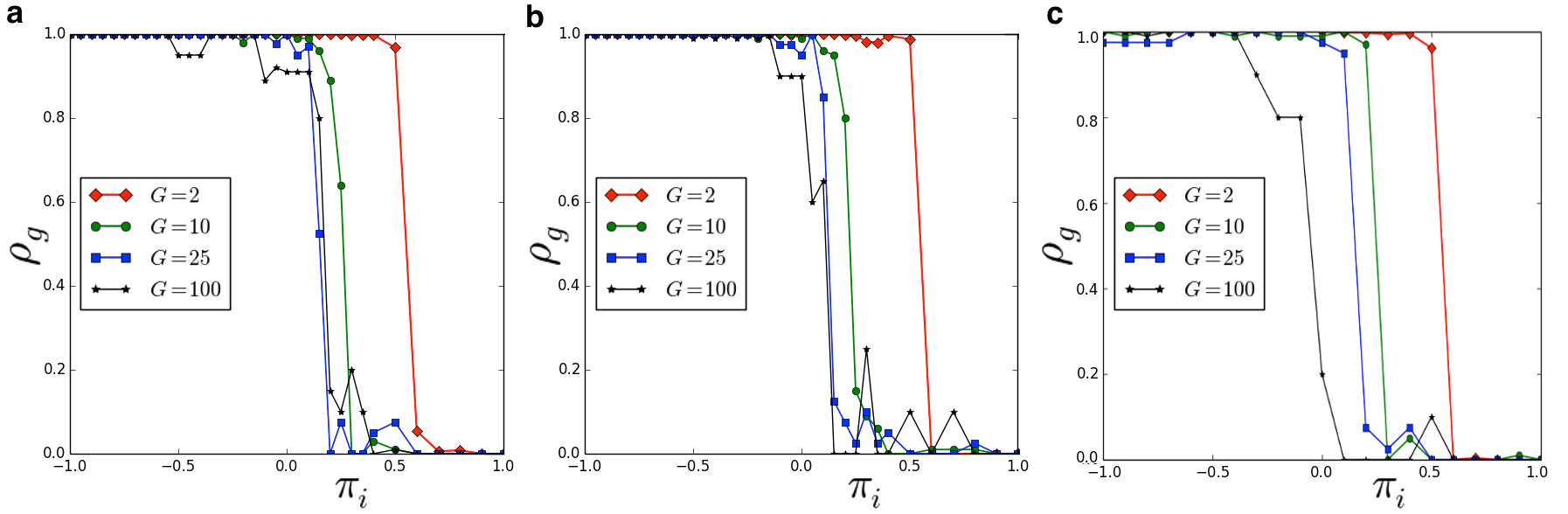}
\caption{Density of groups $\rho_g$ in function of the 'individual payoff' $\pi_i$, on varying the length of the spin vector $L$: \textbf{a} $L = 3$. \textbf{b}$L = 10$. \textbf{c} $L = 25$. Results have been averaged over different simulation runs. \label{fig:figure_2}}
\end{figure*}

\begin{figure*}[!ht]
\centering
\includegraphics[width=16cm]{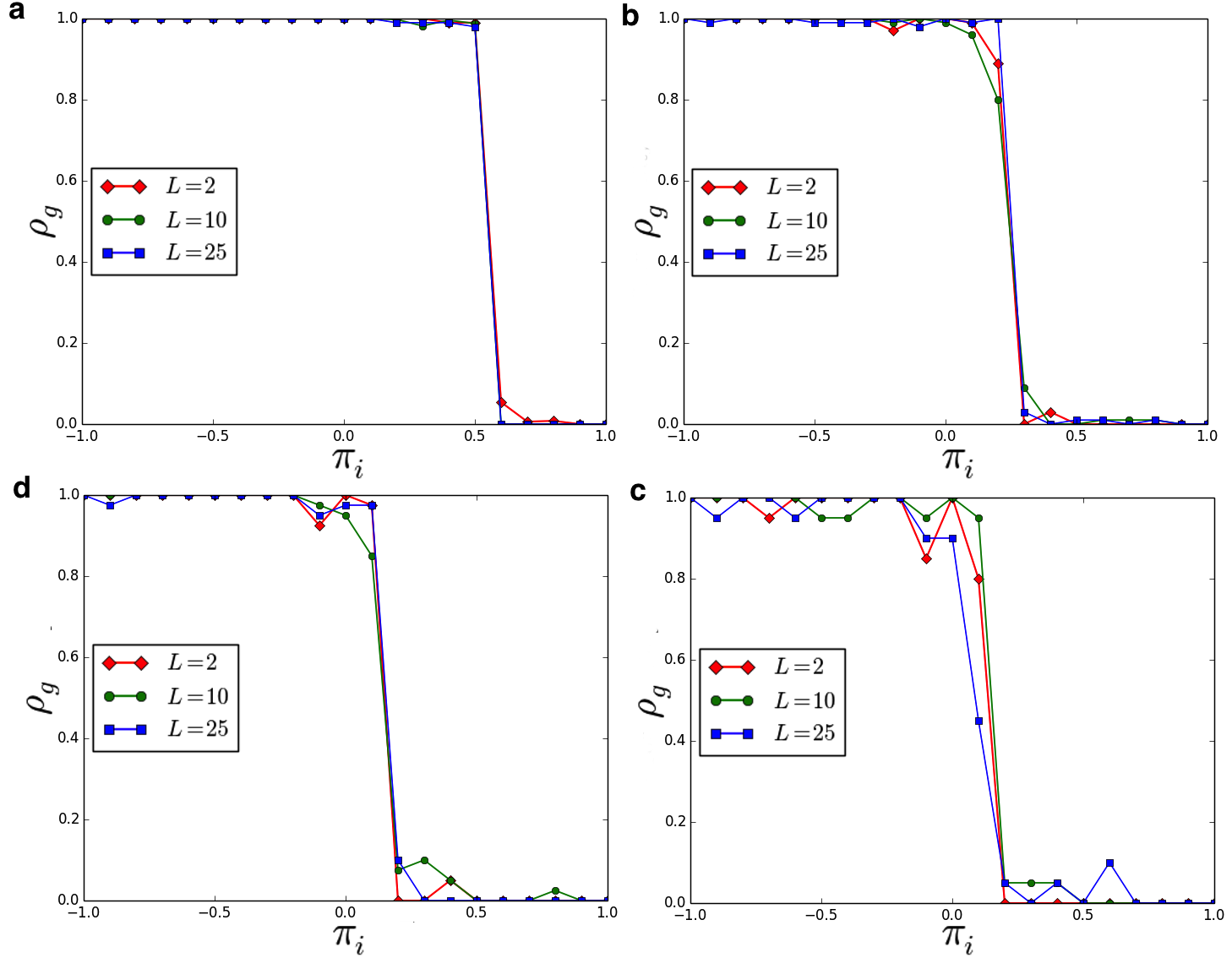}
\caption{Density of groups $\rho_g$ in function of the 'individual payoff' $\pi_i$, for different vector spin length $L$, on varying the group size $G$. \textbf{a}) $G = 2$. \textbf{b}) $G = 10$. \textbf{c}) $G = 25$. \textbf{d}) $G = 50$. Results have been averaged over different simulation runs. \label{fig:figure_3}}
\end{figure*}

\begin{figure*}[!ht]
\centering
\includegraphics[width=16cm]{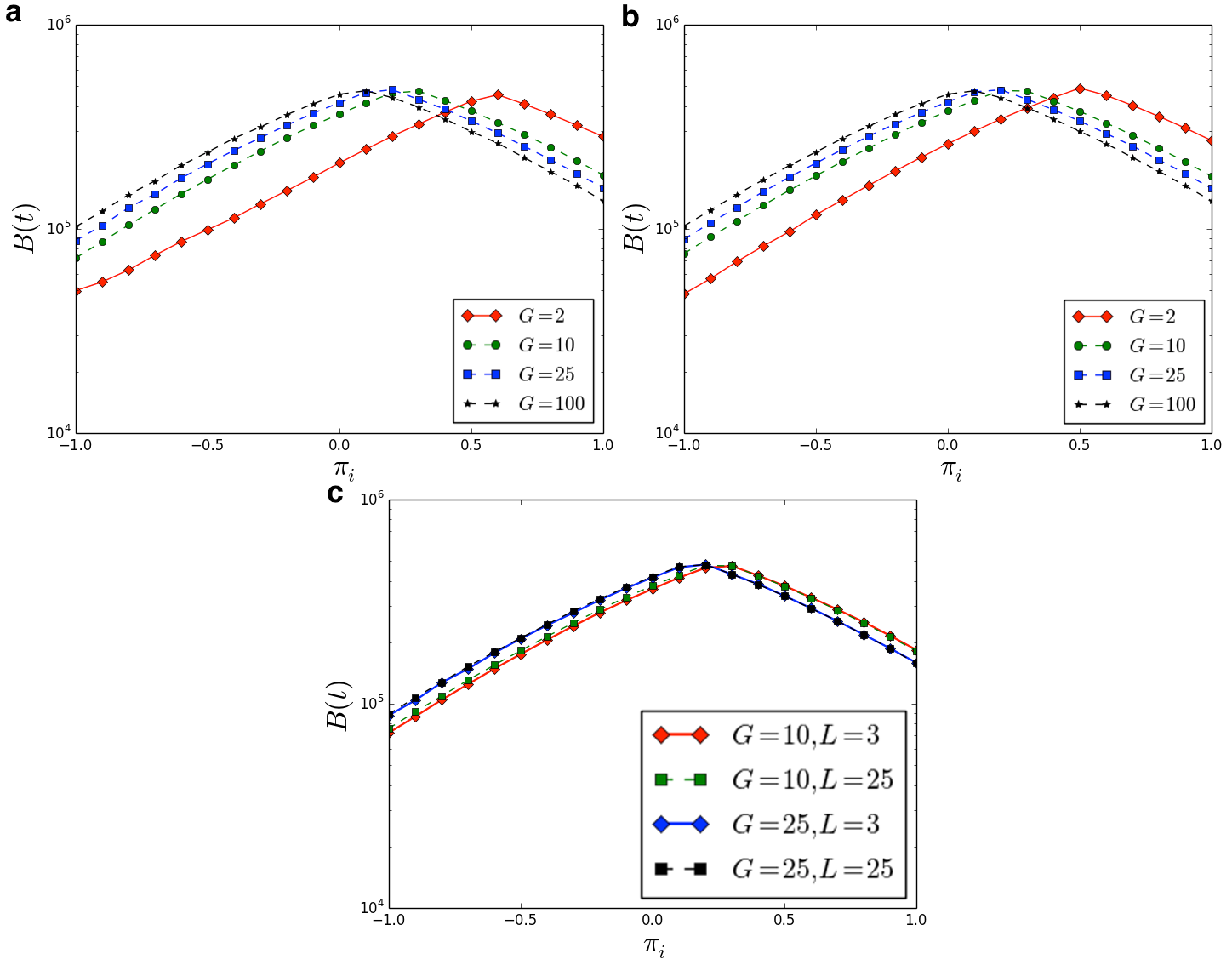}
\caption{Breaking groups over time (i.e. $B(t)$). The legend indicates, for each line, the considered group size $G$. \textbf{a}) Results achieved with $L = 3$. \textbf{b}) Results achieved with $L = 25$. \textbf{c}) Comparison between results achieved with $L = 3$ and $L = 25$. Results have been averaged over different simulation runs. \label{fig:figure_4}}
\end{figure*}

Numerical simulations have been performed in a population with $N = 1000$ agents, considering different conditions related to the 'group payoff' and to the 'individual payoff', i.e. $\pi_i$ in the range $[-1,+1]$, and $\pi_g$ in the range $[0,+1]$.
In addition, we study the dynamics of the population for different length of the spin vector characterizing our agents.
Due to the value of $N$, we analyzed the emergence of groups of the following size: $[2,4,5,10, 25, 50, 100]$.
Figure~\ref{fig:figure_1} shows the phase diagram of our population.
Figure~\ref{fig:figure_2} indicates the density of the groups in function of the 'individual payoff', on varying the length of the spin vector $L$.
It is then possible to find the critical thresholds $\hat{\pi_i}$, on varying the group size $G$. For instance, in the case $L = 3$, we observe $\hat{\pi_i} = 0.55$ for $G = 2$, $\hat{\pi_i} = 0.15$ for $G = 10$, and $\hat{\pi_i} = 0.05$ for $G = 25$. 
It is then worth to evaluate if the length $L$ (i.e. the length of the spin vector) affects the outcomes of the model ---see Figure~\ref{fig:figure_3}.
In particular, one can observe that $L$ does not influence the density of groups at equilibrium.
Eventually, as reported in Figure~\ref{fig:figure_4}, we analyze the number of breaking groups ($B(t)$) over time. In particular, we consider different group sizes $G$, and spin vector lengths $L$, on varying the individual payoff.
\section{Discussion and Conclusion}\label{sec:conclusions}
In this work, we study the phenomenon of group formation using the framework of EGT. In particular, we introduce a simple model where agents evaluate if clustering together, or acting individually, according to a payoff they may receive if acting in group (named 'group payoff'), or individually (named 'individual payoff').
Under the assumption that the 'group payoff' increases while increasing the homogeneity of a group, we study the formation and the breaking of groups.
Even if further investigations would be required in order to evaluate the outcomes on varying the definition of the 'group payoff', we suppose that the achieved results can be considered general enough for envisioning some interesting speculation related to the evolutionary aspects of group formation in nature.
Notably, observing that groups form in species ranging from ants to birds, and from lions to human beings, we support the hypothesis that this process has evolutionary roots~\cite{parrish01,hofmann01}.
In addition, we suggest that the 'individual payoff' is a relevant parameter representing the ensemble of genetic traits, skills, living environments, and even socio-cultural conditions one can observe in real systems. For instance, we hypothesize that being part of a group is more advantageous in a hostile environment than in a relaxed one, as suggested by some theories related to the formation of shoals of fishes.
So, even considering the same species, we can have individuals acting in very small groups and others in big groups. For example, in the modern civilization~\cite{dizard01,tehrani01}, small groups named families are, nowadays, composed of very few members, while tribes living in wilder environments are more copious.
We deem relevant to emphasize that the proposed model suggests the existence of a critical threshold in the 'individual payoff', leading to a sharp transition in the phase diagram (see Figure~\ref{fig:figure_1}), from a 'group phase' achieved for low values of $\pi_i$ to an 'individual phase' achieved for high values of $\pi_g$. Notably, for high values of the critical $\pi_i$ the group formation is scarcely observed.
Here, 'group phase' and 'individual phase' correspond to the two states our population can achieve at equilibrium, i.e. with agents forming groups or acting individually. 
Finally, results reported in figure~\ref{fig:figure_4} confirm previous findings and provide a further detail. In particular, analyzing the average number of breaking groups $B(t)$, we observe that small groups are more robust than big ones, and the maximum number of breaking groups is in correspondence with the critical threshold $\hat \pi_i$. Furthermore, for very high 'individual payoffs' big groups are more robust than small ones (i.e. the opposite of the case with low $\pi_i$).
To conclude, we highlight that the proposed model represents an application of EGT besides its classical domain, providing results that remarkably corroborate the hypothesis that the emergence of groups in animal species has evolutionary roots.
\section*{Acknowledgments}
MAJ wishes to thank National Group of Mathematical Physics (GNFM-INdAM) for supporting his work, and the mobility funds of the Faculty of Psychology and Educational Sciences of Gent University.

\end{document}